\PassOptionsToPackage{prologue,table}{xcolor}
\documentclass[sigconf,prologue,table]{_acm/acmart}
\usepackage{multirow}
\usepackage{tcolorbox}  % 用于创建带边框的文本框
\usepackage{xcolor}  % 颜色支持
\usepackage[prologue,table]{xcolor}  % 允许表格使用颜色
\usepackage{colortbl}  % 支持单元格和整行变色
\usepackage{graphicx}  % 用于插入图片
\usepackage{caption}    % 让 caption 可自定义格式
\usepackage{float}    % 让 [H] 生效
\usepackage{titlesec}
\renewcommand\footnotetextcopyrightpermission[1]{}
\usepackage{enumitem}
\setlist[itemize]{leftmargin=10pt}  % 调整缩进
\setcopyright{none} % to remove the copyright notice
\titleformat{\subsubsection}
  {\normalfont\bfseries} % 标题加粗
  {\thesubsubsection} % 自动编号
  {1em} % 标题与正文间距
  {} % 额外修饰

\titlespacing{\subsubsection}{0pt}{5pt}{3pt} % 控制间距

\AtBeginDocument{%
  \providecommand\BibTeX{{%
    \normalfont B\kern-0.5em{\scshape i\kern-0.25em b}\kern-0.8em\TeX}}}

\acmConference[]{}{}{}
\begin{document}

\title{Goal2Story: A Multi-Agent Fleet based on Privately Enabled sLLMs for Impacting Mapping on Requirements Elicitation}

\author{Xinkai Zou}
\affiliation{
  \institution{Tongji University}
  \state{Shanghai}
  \country{China}
}
\email{jayzou@tongji.edu.cn}

\author{Yan Liu}
\affiliation{
  \institution{Tongji University}
  \state{Shanghai}
  \country{China}
}
\email{yanliu.sse@tongji.edu.cn}
\authornote{Corresponding author.}

\author{Xiongbo Shi}
\affiliation{
  \institution{Tongji University}
  \state{Shanghai}
  \country{China}
}
\email{2151274@tongji.edu.cn}

\author{Chen Yang}
\affiliation{
  \institution{Shanghai Business School}
  \state{Shanghai}
  \country{China}
}
\email{chenyang@sbs.edu.cn}
\renewcommand{\shortauthors}{Zou, et al.}

%%
% Abstract.
\begin{abstract}
As requirements drift with rapid iterations, agile development becomes the dominant paradigm. Goal-driven Requirements Elicitation (RE) is a pivotal yet challenging task in agile project development due to its heavy tangling with adaptive planning and efficient collaboration. Recently, AI agents have shown promising ability in supporting requirements analysis by saving significant time and effort for stakeholders. However, current research mainly focuses on functional RE, and research works have not been reported bridging the long journey from goal to user stories. Moreover, considering the cost of LLM facilities and the need for data and idea protection, privately hosted small-sized LLM should be further utilized in RE. To address these challenges, we propose Goal2Story, a multi-agent fleet that adopts the Impact Mapping (IM) framework while merely using cost-effective sLLMs for goal-driven RE. Moreover, we introduce a StorySeek dataset that contains over 1,000 user stories (USs) with corresponding goals and project context information, as well as the semi-automatic dataset construction method. For evaluation, we proposed two metrics: Factuality Hit Rate (FHR) to measure consistency between the generated USs with the dataset and Quality And Consistency Evaluation (QuACE) to evaluate the quality of the generated USs. Experimental results demonstrate that Goal2Story outperforms the baseline performance of the Super-Agent adopting powerful LLMs, while also showcasing the performance improvements in key metrics brought by CoT and Agent Profile to Goal2Story, as well as its exploration in identifying latent needs.(\url{https://github.com/SoftACE-Lab/goal2story})
\end{abstract}

%    <concept>
%        <concept_id>10011007.10011074.10011075</concept_id>
%        <concept_desc>Software and its engineering~Designing software</concept_desc>
%        <concept_significance>500</concept_significance>
%        </concept>
%    <concept>
%        <concept_id>10010147.10010178</concept_id>
%        <concept_desc>Computing methodologies~Artificial intelligence</concept_desc>
%        <concept_significance>300</concept_significance>
%        </concept>
%    <concept>
%        <concept_id>10002951.10003227</concept_id>
%        <concept_desc>Information systems~Information systems applications</concept_desc>
%        <concept_significance>300</concept_significance>
%        </concept>
%  </ccs2012>

% \ccsdesc[500]{Software and its engineering~Designing software}
% \ccsdesc[300]{Computing methodologies~Artificial intelligence}
% \ccsdesc[300]{Information systems~Information systems applications}

%%
% Keywords.

%% Teaser.


%%
\maketitle
% 右上角插入图片

\begin{figure}[htbp]
    \centering
    \includegraphics[width=0.5\textwidth]{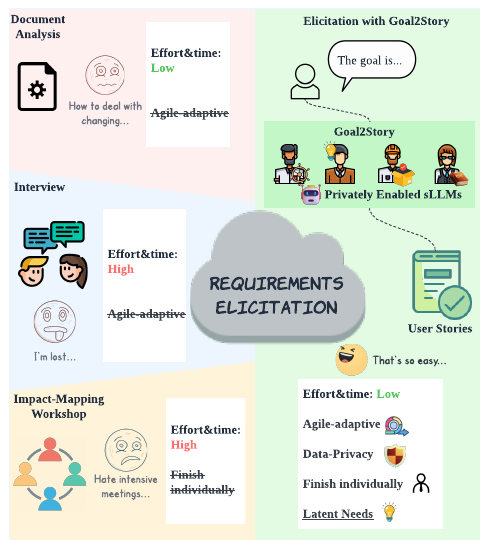}
    \captionsetup{labelfont=normalfont, textfont=normalfont} % 取消加粗
    \caption{\textbf{Inviting Goal2Story:} 
    A Goal-Driven way to mitigate the suffering of practicing requirements elicitation adopting sLLMs, with impact mapping framework}
    
    \label{fig:showcase}
\end{figure}

\section{Introduction}
\label{sec:introduction}

Requirement elicitation(RE) is a pivotal activity in requirements engineering, which extracts and refines stakeholder needs, expectations, and constraints to precisely define system requirements\cite{maalej2013managing}. A well-executed elicitation process minimizes ambiguity, mitigates the risk of requirement drift, and ensures that the software aligns with both business goals\cite{siddeshwar2024comparative} and user demands. Traditional manual-driven RE methods, primarily relying on conventional interviews\cite{ataei2024elicitron} and document-based communication\cite{wei2024requirements},  are inclined to discover user-centric functional needs; while impact mapping (IM)\cite{impactmapping} workshops, grounded in goal-oriented modeling, more effectively aligns features with business goals. With agile development becoming the dominant paradigm, traditional methods like interviews and document-based communication have become less suitable for modern software engineering. The emphasis on rapid iteration, continuous feedback, and adaptive planning in agile processes\cite{choma2016userx} diminishes the effectiveness of these linear, time-consuming and reactive dependency methods, making goal-driven methods like IM a better choice. 

The IM workshop bridges agile strategy and execution, by mapping goals (Why) to user stories (USs)\cite{schon2017agile} through a streamlined hierarchy, which analyzes goal (Why), key actors(Who), relevant impacts(How), and deliverables(What) successively, and from deliverables to USs, ensuring every sprint deliverable directly traces back to measurable business outcomes. While RE is pivotal to product success, human-oriented IM activities often trap teams in subjective assumptions and delayed feedback loops, in addition to causing misalignment between goals and USs, but also exacerbating resource drain in agile environments.

% The LLM-based agent demonstrates remarkable proficiency in handling complex and dynamic software engineering tasks by autonomously perceiving and taking on key roles in various contexts

Recent advances show large language models (LLMs) are gaining attraction in software engineering, with preliminary evidence suggesting LLM-based agents may adapt to complex tasks by autonomously perceiving and taking on key roles in various contexts\cite{liu2024large}. This characteristic gives it the potential to simulate an IM workshop, where the LLM-based agent can act as key roles at different stages of the goal-based hierarchy. The contextual and creative generation capabilities, along with the stable performance of agents, provide the opportunity to generate more adaptable and divergent outputs. This, in turn, enhances the potential to improve the automation of the IM method and increases the alignment between USs and business goals. Additionally, as the central controller of the agent, the LLM, often a commercial model, may be used to process sensitive business data. Security concerns lead many organizations to favor self-hosted or dedicated LLM solutions. However, maintaining private LLMs is financially prohibitive for many organizations\cite{s25051318}\cite{Subramanian2025SmallLM}. Therefore, small-scale LLMs (sLLMs) present a viable trade-off, offering a balance between computational feasibility and privacy.

In this paper, we introduce Goal2Story, a multi-agent fleet intended for goal-driven RE using IM, leveraging private sLLMs as the central agent controller. Goal2Story replicates human workflows derived from IM activities, which facilitates meaningful collaborative interactions among crews through structured outputs. More illustratively, in a fleet rooted in Goal2Story, all crews follow a streamlined workflow to ensure consistency and efficiency. Given a predefined business goal(e.g. increasing daily active users by 5\%), the Alpha Captain is responsible for identifying the most n critical actors, the Intelligence Officer conducts dependency mapping to model how each actor's behavior influence the goals. The Delivery Coordinator then identifies specific impact-enabling features or tasks, which are further decomposed into USs by the Tactical Officer. Since agile projects are influenced by various dynamic factors, agent roles and skill configurations are more complex in practice. Therefore, we further boost the system in terms of two key features: profiles\cite{Wang2023ASO} configured for the agent and Chain-of-Thoughts (CoT)\cite{10.5555/3600270.3602070} reasoning.

To evaluate our approach, we construct StorySeek, and a semi-automatic dataset for agent-based IM activities. StorySeek comprises over 1,000 real-world agile project instances, each containing a business goal, the corresponding USs, project background, and other critical IM elements. Furthermore, we introduce two novel evaluation metrics: (a) Factuality Hit Rate (FHR), assessed through agent-based automated evaluation; and (b) Quality And Consistency Evaluation (QuACE), evaluated through hybrid agent-expert validation.

The experimental results show that Goal2Story consistently delivers strong performance in FHR and QuACE, achieving peak scores of 77.93\% and 97\%, respectively. These results demonstrate that our method can effectively generate high-quality USs, as well as exploring latent needs, even under limited computational resource and strict security constraints. Besides, CoT improves the reasoning efficiency and logical consistency of individual agents, which directly contributes to a 6\% increase in FHR. Agent-specific profiling strategies, though effective in achieving higher performance through amplified contextual relevance and complementarity, while tentatively exhibiting less pronounced enhancements in adaptation to specific project scenarios. Notably, although the selection of different foundation models may influence overall performance, it is the IM framework-powered multi-agent decomposition mechanism that remains the primary factor in optimizing RE outcomes.

Our contributions can be summarized as follows:
\begin{enumerate}
    \item We pioneer the application of multi-agent based requirement elicitation within goal-driven agile processes, significantly enhancing automation capabilities in contemporary software requirements engineering practices.
    \item We propose Goal2Story, a multi-agent fleet for goal-driven agile RE that operates solely with private sLLMs, among which agent powered by LLMs enhances output quality through CoT, while predefined role-specific profiling optimize their domain-aligned communication efficiency.
    \item We introduce a semi-automatic dataset construction methodology to obtain the StorySeek dataset, designed for evaluating goal-driven RE tasks and potential industrial applications.
    \item We define FHR and QuACE, two novel evaluation metrics for assessing goal-driven RE outputs, and present an automated evaluation methodology, with human alignment validation. 
    % Experimental results show that Goal2Story outperforms the baseline performance of Super-Agent adopting powerful LLMs, as well as the impacts from CoT and Profile and latent needs identification. 
\end{enumerate}

% The structure of this paper is organized as follows.

\section{Methodology}
\label{sec:methodology}
% \section{How Goal2Story Functions}
\begin{figure*}[t]  % 使用 figure* 并指定 t (top) 以确保在顶部
  \centering
  \includegraphics[width=\textwidth]{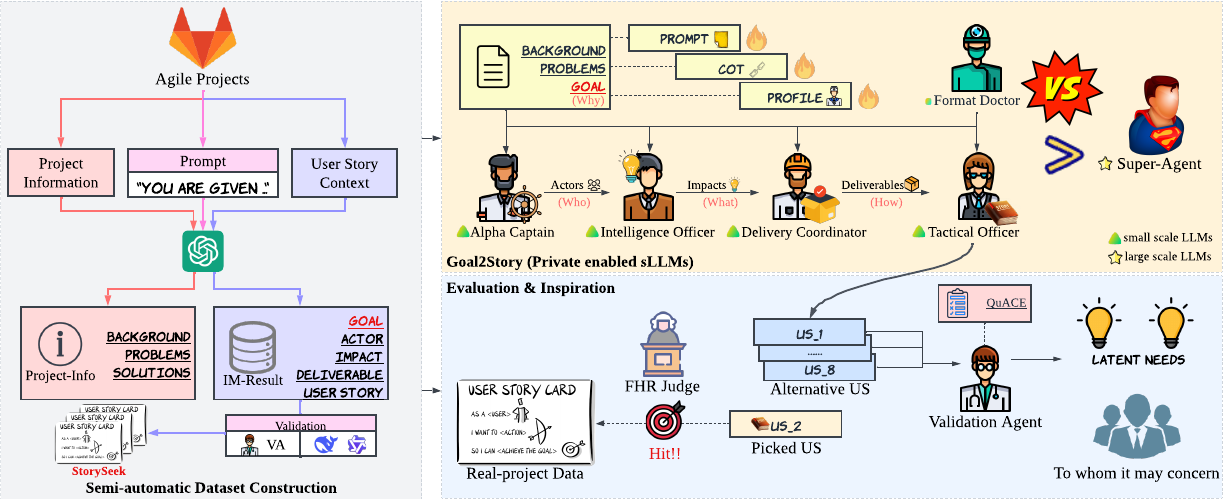}
  \captionsetup{labelfont=normalfont, textfont=normalfont} % 取消加粗
  \caption{\textbf{How Goal2Story Functions.} 
    The left section presents the StorySeek dataset along with the semi-automatic dataset construction method. The upper right section outlines the basic architecture and workflow of Goal2Story, incorporating CoT and Profiling features. The lower right section provides the evaluation results based on FHR and QuACE metrics, as well as insights into latent needs identification.}
  \label{fig:methodology}
\end{figure*}  
\subsection{Research Design}
% 对应整体goal2agent框架，更具有系统性，每个agent更纯粹、各司其职
% 对应难以评价的贡献

Our methodology aims to address two critical challenges in goal-driven RE: 
% (1) Limited research exists on business goal-driven automated RE, and manual processes are time-consuming and susceptible to subjective assumptions, increasing the risks of misalignment of business goals.
% underscoring the imperative to advance both automation efficacy and requirements alignment accuracy.
(1) Current goal-oriented RE processes remain labor-intensive and prone to subjective influence; 
(2) Misalignment between business goals and implemented features due to the delayed feedback from post-implementation validation cycles. 
Building upon multi-agents and IM framework, we propose Goal2Story, a goal-driven approach combining profiling agents and CoT. 
Goal2Story operationalizes the IM activities from agile requirements elicitation, systematically decomposing business goals into a hierarchy of streamlined autonomous agents. Each agent maintains functional purity with explicitly defined boundaries and responsibilities, while leveraging LLMs to generate contextually optimized outputs through advanced reasoning capabilities automatically.
To benchmark the performance of our multi-agent framework with sLLMs as computational kernels under task decomposition paradigms, this study constructs a Super Agent (SA) baseline employing large-scale parameter LLMs that holistically integrate all sequential steps from the IM framework into unified prompt.

Meanwhile, this study addresses the research gap in goal-oriented RE through empirical validation. We constructed a dataset from real-world agile projects and developed quantitative metrics to analyze differences between Goal2Story-generated USs and actual development artifacts. Furthermore, to tackle the persistent challenge of alignment assessment in agile contexts, we propose automated, expert-validated metrics for evaluating business goal-US consistency.

\subsection{Method Description}
\label{sec:overview}
% Architecture Overview

% Core Considerations

% storyseek

% StorySeek
In goal-driven RE, aligning project requirements with business objectives offers significant flexibility and adaptability, particularly within agile development contexts. However, this approach often relies on the availability of well-defined goals and comprehensive project information at the project’s inception—a condition that is difficult to satisfy in practice due to the high cost and time required to gather such data. Besides, the absence of an open-source dataset that encapsulates both user stories and project metadata exacerbates this challenge, hindering the development and evaluation of automated goal-driven RE tools.
To mitigate this gap, a semi-automatic dataset construction method is introduced, resulting in StorySeek. 
This dataset is built from raw data extracted from NEODATASET\cite{Neo2024UserST}, including goals, actors, impacts, deliverables, user stories and corresponding project information were extracted using LLM (GPT-4o) as the processing kernel. Furthermore, the reliability of StorySeek must be rigorously validated.

% Goal2Story
Inspired by the IM framework, we construct Goal2Story as a multi-agent fleet, where each agent, solely relying on sLLMs, is assigned decomposed subtask with hierarchical role. As shown in Figure~\ref{fig:methodology}, Alpha Captain (AC), Intelligence Officer (IO), Delivery Coordinator (DC), and Tactical Officer (TO) are responsible for generating actors, impacts, deliverables, and USs, respectively. Each agent receives the outputs from all preceding agents along with the initial input. To ensure structured and stable outputs, the Format Doctor in the fleet assists each agent in maintaining the correct format. Additionally, we explore the impact of applying Chain-of-Thought (CoT) and Profiling techniques to each agent to assess their effect on performance.

% Evaluation
To evaluate the trustworthiness of Goal2Story, we introduce Factuality Hit Rate (FHR) to quantify the proportion of generated USs that correspond to real project data for each goal. A higher FHR indicates a greater likelihood that Goal2Story produces USs aligned with real-world scenarios, thereby increasing stakeholder confidence. While USs adhering strictly to validation set demonstrate implementation fidelity, such constraint does not inherently indicate defects within the USs themselves. Instead, they may serve as promising alternatives or latent needs for stakeholders, if they are proven reasonable and aligned with the initial goal. Therefore, we also propose Quality And Consistency Evaluation (QuACE) to assess the quality and goal alignment of the generated USs.

\section{Goal2Story: A Multi-Agent Fleet}

\subsection{Architecture and Crews}
Building upon the IM framework's foundations, our Goal2Story architecture implements a multi-agent fleet wherein each sLLM based agent executes specialized decomposed task through predefined role hierarchies.
As shown in Figure~\ref{fig:methodology}, the framework systematically synthesizes initial inputs comprising project context, identified pain points, and defined business goals (the "Why"), mirroring the collaborative dynamics of structured brainstorming sessions in IM workshops.

Goal2Story Fleet includes the following crew members: 
\begin{itemize}
    \item \textbf{Alpha Captain (AC)}: identifies key actors (Who) and selects the n most significant ones.
    \item \textbf{Intelligence Officer (IO)}: identifies potential impacts (What) and selects the n most influential ones for each actor.
    \item \textbf{Delivery Coordinator (DC)}: determines possible deliverables (How) and selects the n most relevant ones for each impact.
    \item \textbf{Tactical Officer (TO)}: generates structured USs based on the initial input and information provided by other agents.
    \item \textbf{Format Doctor (FD)}: validates and refines the JSON output to ensure correct formatting.
\end{itemize}
This modular design allows Goal2Story to systematically break down and process complex requirements while maintaining consistency across generated outputs. In detail, AC, IO, DC, and TO share the same prompt template as shown in the below:
\begin{tcolorbox}[colback=pink!10, colframe=black!50, left=5pt, right=5pt, top=1pt, bottom=1pt, boxrule=0pt, coltitle=black, sharp corners=southwest]
\textbf{Role:} You are an agile requirement expert and good at...
\ 

\textbf{Ref Info:} {background, problems, goal} + {actor} + ...
\ 

\textbf{Task}: Generate the n most ...
\ 

\textbf{Guidelines:} Definition + Writing tips
\ 

% This section of the prompt defines the agent’s area of expertise, which is good at RE and to analyze IM elements. This section is static in the basic setting of Goal2Story while dynamic profiles for each agent can be found in Section~\ref{sec:cot&profiles}
% This section contains static information including background, problems and the goal of the current case, together with dynamic preceding outputs of former agents, such as the actor and impact with only one selection at a time.
% We define four tasks for different agents, in the format: "Generate the two most relevant (one of the IM elements)..."
% We provide a set of guidelines. First, we instruct the agent to output in the format of the given JSON example. Second, based on different tasks, the agent is given specific rules on how to generate the IM element. For example, actions in the US should start with a verb.

\textbf{Output Format:} JSON example
\end{tcolorbox}
\begin{itemize}
    \item \textbf{Role}: The Role component architecturally defines an agent's domain expertise through a dual-layer structure: (1) a static generic context establishing foundational competence (mainly as RE specialists), and (2) a dynamic role instantiation enabling context-specific adaptations through task-dependent parametric configurations, which can be found in Section~\ref{sec:cot&profiles}. 
    \item \textbf{Ref Info}: The Ref Info component integrates both static and dynamic elements: (a) immutable contextual anchors comprising background, problems, and the goal of the current case; (b) operational traces capturing prior agent outputs, such as the actor and impact with only one selection at a time.
    \item \textbf{Task}: The Task component realizes core agent objectives through structured generation of four workflow-aligned intermediate artifacts that map to distinct IM elements. This is implemented via a templated prompt in the format: "Generate the two most relevant [IM element]..."
    \item \textbf{Guidelines}: The Guidelines component prescribes the output format by well-designed JSON examples, and meanwhile the agent is given specific rules on how to generate the IM element. For example, actions in the US should start with a verb.
    \item \textbf{Output Format}: The Output Format component provides the agent with one expected example for reference. 
\end{itemize}
Relatively, the FD setting is much simpler. FD verifies the output format of each agent by calling python \textit{json.loads} tool. If the validation fails, FD will correct the format of original outputs using LLMs with the following prompt:
\begin{tcolorbox}[colback=pink!10, colframe=black!50, left=5pt, right=5pt, top=1pt, bottom=1pt, boxrule=0pt, coltitle=black, sharp corners=southwest]
Please repair the following JSON format, only return the repaired valid JSON content, do not add any explanation or other text, do not use code block tags:  
\{original json content\}
\end{tcolorbox}
\subsection{Workflow}
\begin{figure}[h]
    \centering
    \includegraphics[width=0.45\textwidth]{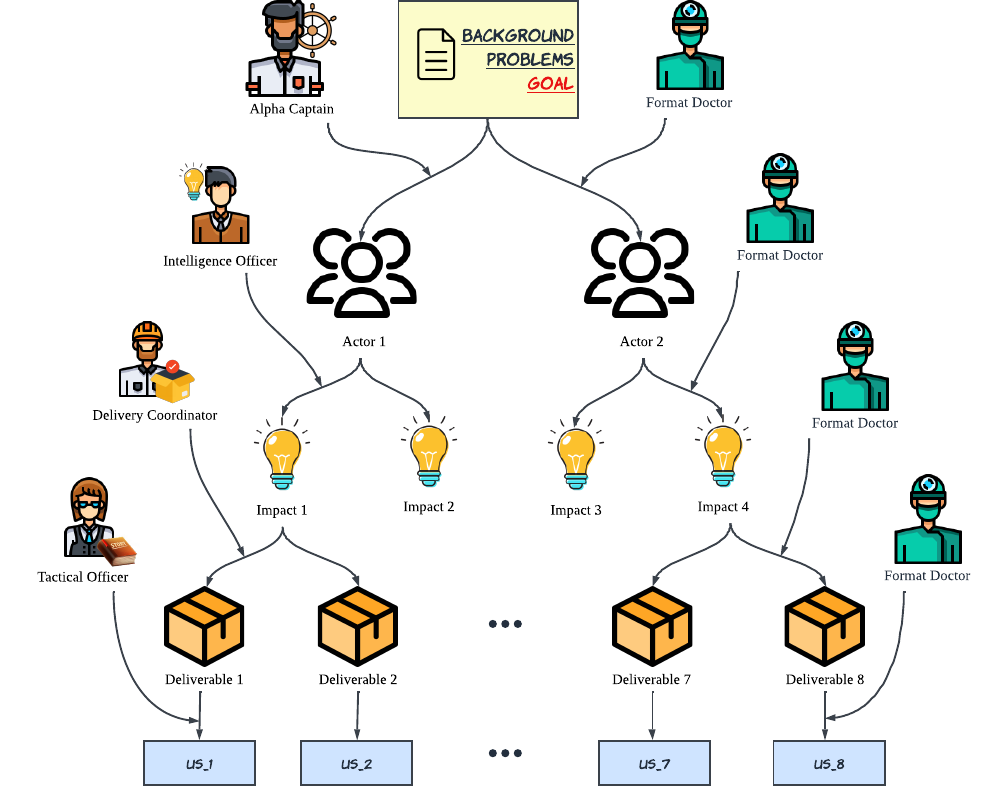}
    \caption{Workflow of the Goal2Story}
    \label{fig:workflow}
\end{figure}
This section provides a detailed introduction to the workflow of Goal2Story. As mentioned earlier, each agent is assigned the task of generating n (default: 2) IM elements at a time. However, to improve the quality of the generated output, after each agent produces its results, the code parses the JSON output and stores it in a temporary array. Then, only one result is sequentially selected at a time, combining it with the outputs of previous agents as the Ref Info for the next agent. This approach reduces the complexity of each task, thereby enhancing the agents' performance. Additionally, to ensure that the output can be correctly processed by the code, the Format Doctor (FD) validates the results and fixes any outputs that do not conform to the expected format. The detailed workflow is illustrated in Figure~\ref{fig:workflow}, where the example demonstrates a case with n=2, ultimately generating eight USs from a single goal.
\subsection{CoT and Profile}
\label{sec:cot&profiles}
To further enhance the performance of Goal2Story, we explore two key techniques:
\begin{itemize}
    \item Chain-of-Thought (CoT): We apply CoT prompting to all agents in both Goal2Story and the SA, allowing them to break down reasoning into step-by-step processes before generating outputs. This improves coherence and accuracy. In Goal2Story, the CoT process can be illustrated as follows:
\begin{tcolorbox}[colback=pink!10, colframe=black!50, left=5pt, right=5pt, top=1pt, bottom=1pt, boxrule=0pt, coltitle=black, sharp corners=southwest]
\textbf{Step 1:} Understanding the given Ref Info...
\ 

\textbf{Step 2:} List all the possible choices...
\ 

\textbf{Step 3:} Validate and prioritize...
\ 

\textbf{Step 4:} Choose 2 most...
\ 
\end{tcolorbox}
    \item Agent Profile: we introduce Profile mechanisms that allow agents to specialize in tasks that best match their designated roles, further improving the quality of generated results. Detailed profile information is provided below:
\begin{tcolorbox}[colback=pink!10, colframe=black!50, left=5pt, right=5pt, top=1pt, bottom=1pt, boxrule=0pt, coltitle=black, sharp corners=southwest]
\textbf{Alpha Captain:} You are an experienced Product Owner who is skilled in identifying key actors...
\ 

\textbf{Intelligence Officer:} You are an experienced Business Analyst who is good at identifying key impacts...
\ 

\textbf{Delivery Coordinator:} You are an experienced Scrum Master who is proficient in analyzing key deliverables...
\ 

\textbf{Tactical Officer:} You are an experienced Product Owner who is good at identifying key user stories...
\ 
\end{tcolorbox}
\end{itemize}

\section{StorySeek Dataset}
\subsection{Details of Dataset Construction Method}
Section~\ref{sec:overview} provides an overview of the semi-automatic dataset construction method based on real-world Agile projects hosted on GitLab. Specifically, as the first step, we selected 10 representative GitLab projects that adopt Agile methodologies from NEODATASET, due to its timeliness and the fact that the data is directly extracted from GitLab issue boards, ensuring strong consistency and strict alignment with real-world scenarios. These 10 projects encompass software projects from various domains, enhancing the diversity of the dataset. The raw data underwent preprocessing, where records with insufficient information were filtered out, and USs containing 10 or fewer words of context were removed. Then, we employed the GPT-4o\footnote{\url{https://openai.com/index/gpt-4o-system-card/}} model to generate IM-Result and Project-Info, setting the temperature to 0.3 and enforcing a structured JSON format using OpenAI's custom structured output tool. Finally, these structured outputs were extracted from the JSON format to construct the StorySeek dataset.

\subsection{Details of StorySeek}
The StorySeek dataset comprises 1,005 records, each containing parsed information from detailed IM-Result and Project-Info, as previously described. Each datapoint includes both the IM-Result and project information. Specifically, IM-Result consists of five key components: goal, actor, impact, deliverable, and US, where the US explicitly details the actor, action, and expected outcome. The project information includes background, problems, and solutions. The expected contribution of this dataset is to support research and industry applications in RE. Although StorySeek was initially designed for evaluating our Goal2Story, it also contains elements relevant to other aspects of software engineering, making it possible to explore new findings in future studies. Our hugging-face repository presents the basic information of the selected GitLab projects along with an example from the dataset. (\url{https://huggingface.co/datasets/SoftACE/StorySeek})

\subsection{Verification of StorySeek}
\label{ref:veri_ss}
To maintain the quality and stability of the StorySeek dataset, we carried out two verification experiments. 
\ 

(1) Quality Assessment by the Validation Agent: We used an automated validation process with a Validation Agent (VA) based on large-scale LLMs, utilizing models distinct from GPT-4o. The VA was designed to evaluate high-quality USs and perform fact-checking to ensure dataset reliability using the QuACE framework, and more details about QuACE can be found in Section~\ref{sec:quace_detail}. The validation process involved the generated US and goal, along with the original US context, forming a new data pair that was then input into the VA for assessment. Specifically, we selected all records from the StorySeek and used GPT-4o-mini\footnote{\url{https://openai.com/index/gpt-4o-mini-advancing-cost-efficient-intelligence/}} and Qwen2.5-72B-Instruct\footnote{\url{https://huggingface.co/Qwen/Qwen2.5-72B-Instruct}} as the base LLMs for the VA. If the generated US met all QuACE criteria and was deemed high quality, the VA assigned a score of "1"; otherwise, it assigned a "0". The verification results showed that StorySeek achieved a high-quality rate of 98.01\% using GPT-4o-mini and 99.90\% using Qwen2.5-72B-Instruct, with an alignment rate of 97.91\% between these two models. This indicates that our dataset maintains high quality and that a single LLM can be effectively used for the VA in QuACE-based evaluation of generated USs for further Goal2Story evaluation.
\ 

(2) Cross-Model Stability Validation: Given concerns about the closed-source nature of certain LLMs\cite{Tao2024ImgTrojanJV}, we conducted a comparative analysis by generating user stories using both open-source and closed-source LLMs. The consistency of these outputs was further verified through human evaluation. We checked content consistency, considering that ChatGPT is a closed-source model. To do this, we randomly selected 15 records from the original dataset and generated IM-Result and Project-Info using DeepSeek-R1 and Qwen2.5-72B-Instruct for comparison. A human alignment team was formed to evaluate whether the outputs from different models were logically and factually consistent. Specifically, the team consisted of one Agile coach (Judge A), one Software Engineering undergraduate (Judge B), and one individual with no relevant experience (Judge C). The team was provided with a basic task description before the task, and then finished the alignment test separately. The validation results indicate that 80\% of the cases met the criterion of "at least two user stories being consistent," while over 37\% were classified as "three user stories being consistent." These findings suggest that StorySeek maintains both logical and factual consistency. A more detailed analysis is provided in Appendix~\ref{app:alignment}.

\section{Experiments}
\subsection{Settings}
This section describes the implementation details of Goal2Story and Super-Agent. To evaluate Goal2Story, we conducted experiments using three open-source, small-scale LLMs: Llama-2-7B\footnote{\url{https://huggingface.co/meta-llama/Llama-2-7b}}, Qwen2.5-7B-Instruct-1M\footnote{\url{https://huggingface.co/Qwen/Qwen2.5-7B-Instruct-1M}}, and DeepSeek-R1-Distill-Qwen-7B\footnote{\url{https://huggingface.co/deepseek-ai/DeepSeek-R1-Distill-Qwen-7B}}. These models were deployed locally using OLLAMA on MacBook Pro with M2 Pro chips and memory of 16GB to simulate real-world scenarios. Regarding parameter settings, AC generates two actors for the given goal, while IO and DC generate two impacts per actor and two deliverables per impact. Finally, TO produces one US that aligns with the preceding results, so that 8 USs are generated at one iteration. As no standardized benchmarks exist for evaluating goal-driven RE automation, we establish a Super-Agent (SA) as a baseline. The SA employs a powerful, large-scale LLM to directly integrate all steps of the IM framework into a single-prompt inference, using CoT technique. For comparison, we tested Super-Agent using GPT-4o and DeepSeek-R1\footnote{\url{https://huggingface.co/deepseek-ai/DeepSeek-R1}}, as they offer strong reasoning capabilities. These two models were evaluated through API-based inference on their development platform. To further analyze our results, we conducted additional experiments on Goal2Story with and without CoT and Profiling features. Notably, during preliminary experiments, we found that Super-Agent without CoT often generated low-quality outputs, even when using large-scale reasoning LLMs such as DeepSeek-R1, which was not robust enough to serve as a baseline. Therefore, we only included Super-Agent with CoT as the baseline in our experiments.
\subsection{Evaluation Metrics}
\subsubsection{Factuality Hit Rate}
\noindent To evaluate the trustworthiness of the Goal2Story, we introduce Factuality Hit Rate (FHR) as a metric to quantify the proportion of generated USs that correspond to real project data for every goal. A higher FHR indicates a greater likelihood that the Goal2Story produces user story predictions aligned with real-world scenarios, thereby enhancing stakeholder confidence. To compute FHR automatically, we employ a text embedding model to measure the similarity between elements in the generated US and those in the dataset. Given a set \( S_G = \{US_{g1}, US_{g2}, ..., US_{gn}\} \) of \( n \) user stories generated by Goal2Story for a specific goal \( G \), and the US \(US_{g}\) in the dataset associated with the same goal, we define a matching criterion based on element-wise similarity. Each US consists of three elements including actor, action and expected outcome, denoted as \( E \). If at least one generated user story \( US_i \) satisfies the similarity threshold \( T_e \) for every element \( e \) that is validated via human alignment, with respect to a real user story \( US_{g} \), the generated user story is considered a "hit", otherwise is a "no-hit":

\begin{equation}
sim(US_{gi}^e, US_{g}^e) \geq T_e, \quad \forall e \in E
\end{equation}

where \( US_i^e \) and \( US_{g}^e \) represent the textual representations of element \( e \) in the generated and the dataset USs, respectively, and \( sim(\cdot, \cdot) \) denotes the similarity function based on the text embedding model. If at least one generated US \( US_i \) meets this criterion for any real US \( US_{g} \), we consider the real US as hit by Goal2Story.  

To compute the FHR, we define \( N_{\text{hit}} \) as the number of goals for which at least one generated US hits a real US in the dataset, and \( N_{\text{total}} \) as the total number of goals in the dataset. The FHR is then computed as:

\begin{equation}
FHR = \frac{N_{\text{hit}}}{N_{\text{total}}}
\end{equation}

To evaluate, we employed a text embedding model to measure the similarity between each element in the generated US and those in the dataset. Specifically, we used text-embedding-v3\footnote{\url{https://help.aliyun.com/zh/model-studio/user-guide/embedding}} from Qwen and manually set similarity thresholds at 0.7 for the actor, 0.6 for the action, and 0.6 for the expected outcome. These thresholds were validated through human alignment experiment. Based on the similarity results, we further performed a statistical analysis using the FHR metric to evaluate different methods across various projects.

\subsubsection{Quality And Consistency Evaluation}
\label{sec:quace_detail}
\noindent In addition to FHR, we observe that USs generated beyond those in the dataset can also be valuable, provided they exhibit logical coherence and high quality. Such USs serve as alternative pathways to achieving project goals. To evaluate their quality, we propose the Quality and Consistency Evaluation (QuACE) metrics, which integrate selected criteria from traditional Quality User Story (QUS) metrics \cite{10.1007/s00766-016-0250-x} with goal-related and factual consistency measures. QuACE consists of four major evaluation dimensions: Syntactic, Semantic, Pragmatic, and Consistency.

Previous studies \cite{Ronanki2023ChatGPTAA} have demonstrated that employing LLMs for US quality assessment achieves high alignment with human evaluation. Inspired by this, we propose a Validation Agent (VA) that utilizes LLMs to apply the QuACE framework through carefully designed prompts. In this paper, QuACE is used to assess both the validity of the StorySeek dataset and the quality of generated USs. Specifically, only USs that meet all QuACE criteria are considered high quality, as they effectively explore latent needs in RE, offering alternative solutions that stakeholders might not have initially considered.

Following the VA approach, we assigned it the task of assessing US quality using the QuACE framework. In particular, the Consistency dimension, integrated into the traditional QUS framework, measures the alignment of generated USs with the goal. We selected Qwen2.5-72B-Instruct as the base LLM for the VA. For evaluation, we randomly selected 100 generated USs that extend beyond the original dataset, drawn from both Goal2Story and Super-Agent under different model and technique configurations. These USs, along with project information, were passed to the VA, which classified them as "1" (high quality) or "0" (low quality), recording failed criteria names. This process allowed us to compute the proportion of high-quality USs among all generated USs beyond the dataset. This metric serves as a key indicator of how effectively the system expands the unknown-unknown knowledge space \cite{Jiang2024IntoTU} in RE, uncovering latent needs that stakeholders might not have initially considered. Further details about QuACE are provided in Appendix~\ref{app:QuACE_app}.

\subsection{Human Alignment Check}
\noindent Additionally, we evaluate two methods designed for the automatic assessment of FHR and QuACE through human alignment experiments. We use F1 score, Alignment Rate, and False Positive Rate (FPR) as key metrics to assess the human alignment process for these two annotation methods. In detail, we grouped an human alignment team and conducted several experiments via different questionnaires with specific tasks, with basic descriptions and instructions of tasks provided. To ensure diverse evaluation perspectives, we assembled an alignment team consisting of an Agile coach, an undergraduate student in Software Engineering, and an individual with no relevant experience.

\begin{itemize}
    \item We randomly selected 10 cases each from both the "hit" and "no-hit" circumstances, and the case contains data of the generated user story and the user story from the dataset. Then we asked every people in the human alignment team to determine whether the case is "hit" or "no-hit" separately with no annotation answer provided.
    \item We also randomly selected 20 cases from the generated user stories of QuACE check. Then, we organized the data with information of background, problems, the goal and the user story, together with QuACE evaluation criteria. The people in the human alignment team determined whether the case is high-quality or not separately. After these two experiments, we calculate the F1, Alignment Rate and FPR for each annotation method. Specifically, Alignment Rate is the proportions of the aligned cases in the total cases.
\end{itemize}

The team was provided with basic instructions and project reference information, and then finished the experiment separately. The validation results for the FHR metric show that F1 score, alignment rate, and FPR achieved 75.41\%, 75\%, and 24.14\%, respectively, demonstrating strong performance. These findings indicate that our approach, which integrates a text embedding model to automate FHR computation, is both effective and well-aligned with human evaluation. For the QuACE metric, the F1 score, alignment rate, and FPR were 75\%, 63.33\%, and 70.59\%, respectively. While the FPR was relatively high, the F1 score and alignment rate remained satisfactory. Further analysis revealed that the model tends to apply stricter syntactic criteria when evaluating USs, whereas in other dimensions, such as Consistency, its assessments closely align with human judgment. This suggests that our automated QuACE evaluation method for generated USs is both reasonable and effective, with further improvement on syntactic aspect of the generated USs. More detailed experimental results can be found in Appendix~\ref{app:alignment}
\section{Findings}
\label{sec:findings}
% 双栏顶部放置表格
\begin{table*}[t]
    \centering
    \renewcommand{\arraystretch}{1.5}  % 增加行间距
    \resizebox{\textwidth}{!}{
    \begin{tabular}{l|l|ccccccccccc|c}
        \hline
        \multirow{2}{*}{\textbf{Method}} & \multirow{2}{*}{\textbf{Model}} & \multicolumn{11}{c|}{\textbf{FHR}} & \textbf{QuACE} \\
        \cline{3-14} 
        & & \textbf{1} & \textbf{2} & \textbf{3} & \textbf{4} & \textbf{5} & \textbf{6} & \textbf{7} & \textbf{8} & \textbf{9} & \textbf{10} & \textbf{Mean} & \textbf{Random100} \\
        \hline
        \multirow{2}{*}{Super-Agent (with CoT)} & GPT-4o & 66.67\% & 53.85\% & 51.52\% & 78.43\% & 68.27\% & 66.96\% & 73.57\% & 62.50\% & 64.42\% & 60.84\% & 64.70\% & 91\% \\
        & DeepSeek-R1 & 33.33\% & 46.15\% & 54.55\% & 48.04\% & 62.50\% & 42.86\% & 62.14\% & 36.25\% & 40.49\% & 30.72\% & 45.70\% & 75\% \\
        \hline
        \multirow{3}{*}{Goal2Story} & Qwen2.5-7B-Instruct-1M & 66.67\% & 84.62\% &\textbf{ 87.88\%} & 81.37\% & 82.69\% & 74.11\% & 86.43\% & 72.50\% & 55.83\% & 72.89\% & 76.50\% & \textbf{97\%} \\
        & Llama2-7b & 75.00\% & 76.92\% & 84.85\% & 77.23\% & 81.55\% & 77.27\% & 85.00\% & 79.87\% & \textbf{68.10\%} & 73.49\% & 77.93\% & 54\% \\
        & DeepSeek-R1-Distill-Qwen-7B & 50.00\% & 46.15\% & 75.76\% & 68.63\% & 61.17\% & 58.04\% & 75.71\% & 54.38\% & 55.83\% & 56.02\% & 60.17\% & 46\% \\
        \hline
        \multirow{2}{*}{Goal2Story (with CoT)} & Qwen2.5-7B-Instruct-1M &\textbf{ 83.33\%} & \textbf{100.00}\% & 72.73\% & 88.24\% & 86.54\% & 76.79\% & \textbf{92.14\%} & 80.00\% & 67.48\% & 74.70\% & \textbf{82.19\%} & 94\% \\
        & Llama2-7b & 75.00\% & 69.23\% & 78.79\% & \textbf{90.20\%} & 85.29\% & \textbf{79.82\%} & \textbf{92.14\%} &\textbf{ 83.33\%} & 65.84\% & 75.15\% & 79.48\% & 61\% \\
        \hline
        \multirow{2}{*}{Goal2Story (with Profile)} & Qwen2.5-7B-Instruct-1M & 75.00\% & 61.54\% & 78.79\% & 81.37\% & \textbf{88.00\%} & 77.68\% & 87.86\% & 77.50\% & 65.64\% & \textbf{77.11}\% & 77.05\% & 92\% \\
        & Llama2-7b & 81.82\% & 76.92\% & 78.79\% & 84.31\% & 86.54\% & 79.46\% & 88.49\% & 81.25\% & 61.35\% & 76.51\% & 79.54\% & 54\% \\
        \hline
    \end{tabular}
    }
    \captionsetup{labelfont=normalfont, textfont=normalfont} % 取消加粗
    \caption{\textbf{Different Methods with Different Models Performance on FHR and QuACE.} This table shows performance of different method-model pair on FHR in every project dataset, as well as mean FHR and QuACE on the whole StorySeek dataset. Random100 indicates the proportion of generated USs that pass QuACE, based on a random selection of 100 generated USs from different method-model pairs.}
    \label{tab:performance}
\end{table*}

\subsection{Goal2Story vs. Super-Agent}
\begin{figure}[h]
    \centering
    \includegraphics[width=0.45\textwidth]{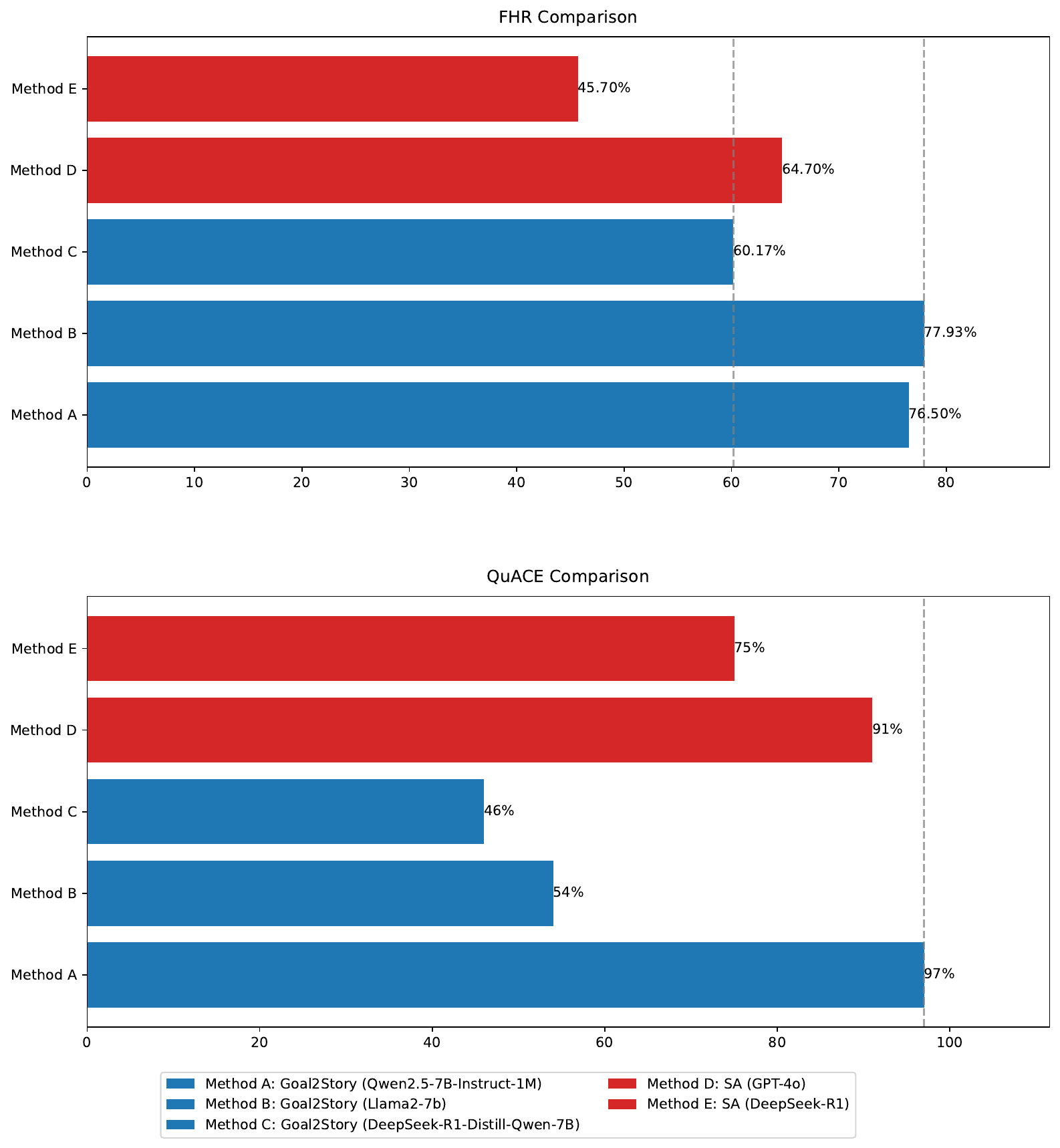}
    \caption{Performance of Goal2Story and Super-Agent (with CoT) with different models on FHR and QuACE }
    \label{fig:gvss}
\end{figure}

As shown in Figure~\ref{fig:gvss}, experimental results reveal significant differences in Factual Hit Rate (FHR) and Quality And Consistency Evaluation (QuACE) between the evaluated methods. FHR quantifies the similarity between generated USs and those in the dataset, while QuACE assesses US quality and goal alignment, particularly in extrapolated cases beyond the original dataset. Results indicate that, despite performance variations across different models, Goal2Story, utilizing only sLLMs, consistently outperforms the Super-Agent with CoT, which relies on a more powerful LLM, in both FHR and QuACE. Notably, in one particular model, Goal2Story demonstrated exceptionally strong performance. Specifically, Goal2Story paired with Qwen-2.5-7B-Instruct-1M and LLaMA2-7B achieves 76.50\% and 77.93\% FHR, respectively, surpassing the 64.70\% FHR of Super-Agent with GPT-4o. Regarding QuACE, Goal2Story with Qwen-2.5-7B-Instruct-1M attains 97\% high-quality US, exceeding Super-Agent (with CoT) using GPT-4o (91\%). Additionally, while Goal2Story with DeepSeek-Distill-Qwen-7B exhibits a slightly lower FHR (60.17\%) than Super-Agent with GPT-4o (64.70\%), it remains superior to Super-Agent with DeepSeek-R1 (45.70\%).

Overall, Goal2Story demonstrates stronger performance despite variations in QuACE when combined with LLaMA2-7B and DeepSeek-Distill-Qwen-7B. Notably, it maintains higher FHR scores, ensuring better alignment with real requirements while enhancing security and cost-efficiency. This approach facilitates structured US generation through goal decomposition and role distribution, effectively reducing deviations from actual user needs.

\subsection{Impact of CoT and Profile}
\begin{figure}[h]
    \centering
    \includegraphics[width=0.45\textwidth]{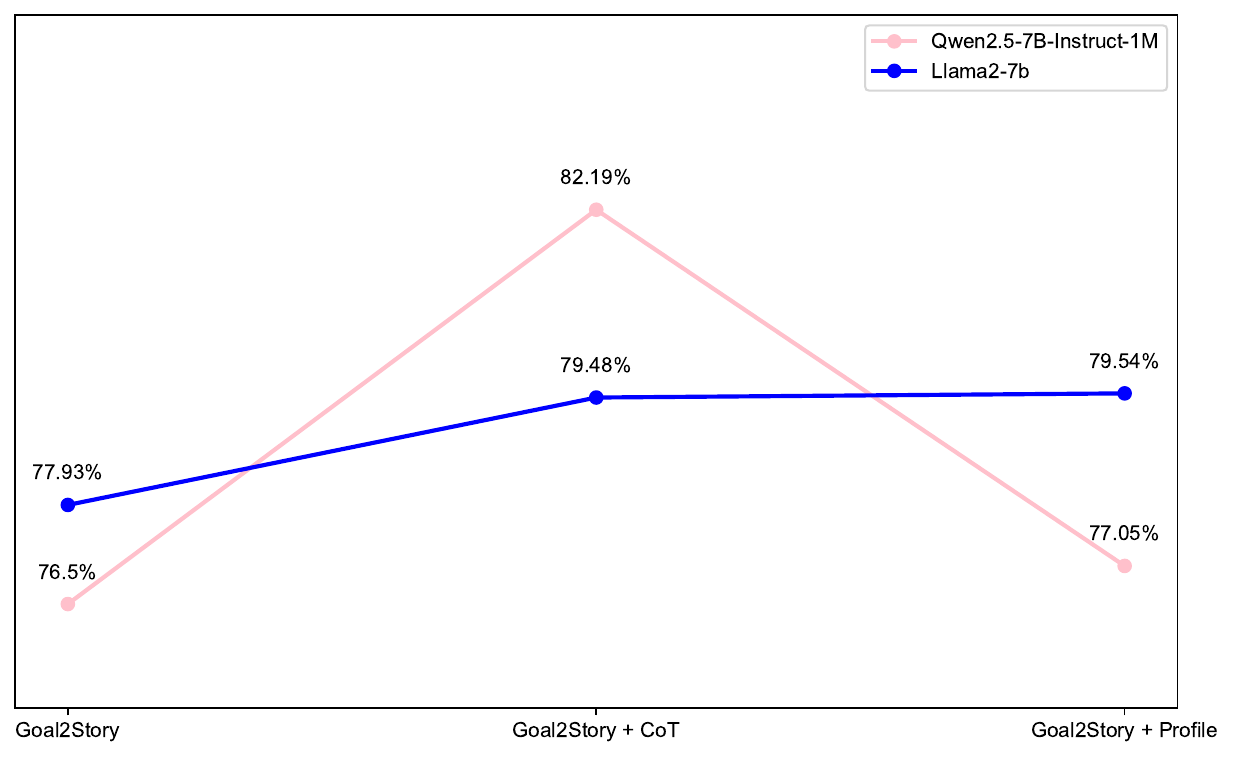}
    \caption{Performance of Goal2Story with different features (CoT and Profile) on FHR}
    \label{fig:feature}
\end{figure}
As shown in Figure~\ref{fig:feature}, the Chain of Thought (CoT) reasoning significantly enhances FHR, while the Profile feature provides improvements in specific settings. When applied to Goal2Story, CoT and Profile yield 82.19\% and 77.05\% FHR, respectively, on Qwen-2.5-7B-Instruct-1M, and 79.48\% and 79.54\% FHR on LLaMA2-7B, consistently outperforming the baseline Goal2Story configuration. The most notable improvement is observed with CoT on Qwen-2.5-7B-Instruct-1M. For QuACE, CoT and Profile also lead to improvements, with Qwen-2.5-7B-Instruct-1M achieving 94\% and 92\%, respectively. These findings suggest that Profile improves FHR performance on alignment with real projects in the dataset, but may have constraints on QuACE for generating high-quality USs.

\subsection{Performance Variability Across Models}
The analysis of different models highlights notable performance trade-offs. As shown in Table~\ref{tab:performance}, instruction-tuned models\cite{Zhang2023InstructionTF} such as Qwen-2.5-Instruct series effectively balance computational efficiency with task alignment, excelling in multi-agent systems where goal decomposition and task distribution are crucial. Among these, Qwen-2.5-7B-Instruct-1M demonstrates superior QuACE performance, maintaining stability in both USs quality and latent needs identification, particularly when extrapolating beyond the dataset. In contrast, open-source base models like LLaMA2-7B perform well in broad language tasks but exhibit greater variability in QuACE when generating high-quality USs, necessitating further prompt optimization or targeted fine-tuning. Meanwhile, distilled models\cite{hsieh2023distilling} such as DeepSeek-Distill-Qwen-7B offer advantages in resource-constrained or domain-specific applications but struggle with complex goal decomposition and requirement expansion.

Ultimately, model size and type are not the sole determinants of RE performance. Instead, effective structuring of generation and evaluation using IM and multi-agent methodologies, complemented by CoT reasoning and Profile configurations, plays a more significant role in ensuring alignment with goal-driven RE. These strategies enable both instruction-tuned and open-source models to achieve strong results, sometimes even surpassing more powerful single-model architectures.

\subsection{Potentials of Goal2Story}
Beyond the primary findings, further analysis revealed additional noteworthy insights.
\begin{itemize}
    \item Diverse IM Pathways Leading to Identical USs: The analysis of Goal2Story outputs reveals that identical USs can emerge from different IM reasoning processes, indicating that multiple cognitive pathways can lead to the same requirement formulation. This phenomenon closely mirrors real-world scenarios, where individuals with different perspectives and reasoning approaches may, seemingly by coincidence, arrive at similar conclusions. This further supports the trustworthiness of the generated USs, as it demonstrates their alignment with diverse yet valid reasoning processes.
    \item Latent Needs Identification: The generated USs not only align with existing data but also uncover hidden requirements, as validated by QuACE. These USs may represent previously undiscovered needs, helping stakeholders consider requirements more comprehensively. Moreover, this effect could be enhanced by increasing the number of generated USs per iteration or broadening the scope of exploration during every tasks assigned to each agent. This suggests that the approach facilitates eliciting stakeholder needs that may not have been explicitly articulated.
\end{itemize}

\begin{tcolorbox}[colback=gray!10, colframe=black!50, left=5pt, right=5pt, top=5pt, bottom=5pt, boxrule=0pt, coltitle=black, sharp corners=southwest]
\textbf{Insights:} Goal-driven RE is inherently suited for agile development, enabling rapid adaptation to evolving user needs. Goal2Story consistently achieves strong results in FHR and QuACE, demonstrating that sLLM-based methods can generate and evaluate high-quality US even under constrained computational and security conditions. Additionally, CoT enhances trustworthiness and high-quality in complex goal decomposition and US generation, while Profile can also improve alignment with real-world data. However, the impact of Profile varies based on the balance between project-specific contextual alignment and generating high-quality USs. Lastly, while model characteristics influence performance, the effectiveness of Goal2Story adopting IM framework remains the dominant factor in optimizing RE outcomes.
\end{tcolorbox}
\section{Related Work}

\subsection{Automatic requirement elicitation}
% 一行需求
% % 需求激励；
% Wei\cite{wei2024requirements} requirement are all you need;
% focus on the one-line requirement

Requirements elicitation(RE), a critical activity in requirements engineering, significantly influences the success of software projects\cite{khanlarge}. Since the emergence of LLMs, numerous studies have investigated their potential applications in RE. Early approaches, such as GPT-eningeer, focused on leveraging LLMs to generate code from highly abstract requirement descriptions\footnote{GPT-Synthesizer: https://github.com/RoboCoachTechnologies/GPT-Synthesizer}. However, Wei et al.\cite{wei2024requirements} advised against overreliance on "single-line requirement" methods in LLM-driven software development, emphasizing the necessity of refining detailed and high-quality requirements.

Interactive elicitation methods like interviews, questionnaires have also been adapted across domains. For instance, Gorer et al.\cite{gorer2023generating} employed LLMs to generate interview scripts for RE, while White et al.\cite{white2024chatgpt} developed customized prompts to articulate system requirements and identify gaps. Beyond textual approaches, Nakagawa et al.\cite{nakagawa2023mape} integrated LLMs with traditional modeling languages, demonstrating their capability to generate goal-oriented models from contextual inputs. Chen et al.\cite{chen2023use} explored GPT-4’s current knowledge and proficiency in specific modeling languages, while Siddeshwar et al.\cite{siddeshwar2024comparative} investigated automated goal model generation from user stories. Notably, Belzner et al.\cite{belzner2023large} highlighted challenges in verifying requirements alignment with project goals.

In agile development, researchers increasingly focus on AI-assisted RE. Marczak-Czajka et al.\cite{marczak2023using} utilized ChatGPT to generate user stories with human-centric values, inspiring novel requirements. Similarly, Zhang et al.\cite{zhang2024llm} applied GPT models to evaluate and refine user story quality. A growing trend involves integrating AI agents into the elicitation process to enhance efficiency and coverage\cite{ataei2024elicitron}.

% 交互式的方法
% 交互式访谈脚本 \cite{gorer2023generating}
% Gorer et al. [19] used LLMs for generating requirements elicitation
% interview scripts, demonstrating the model’s efficacy in enhancing
% the quality of these scripts.
% Belzner\cite{belzner2023large}

% 需求诱导不局限于软件工程

% 敏捷方法
% cabrero2024exploring 敏捷开发 utilization of GPT-4 as assistants in agile software development meetings

% % 近年来，面向goal的方法；建模语言-
% The extent to which such technologies can help with requirements engineering activities, especially the ones surrounding modeling, however, remains to be seen.
% a focus on the development of goal-oriented models

% 从USs中自动生成Goal model
 % In this paper we introduce a technique that leverages Large Language Models (LLMs) to automatically generate goal models from user stories. \cite{siddeshwar2024comparative}
 % User stories, expressed in snippets of natural language text, are commonly used to elicit stakeholder's needs in agile software development. 

% 探索构建 goal models chen2023use
% explore GPT-4's current knowledge and mastering of a specific modeling language, namely the Goal-oriented Requirement Language (GRL).

\subsection{LLM for software engineering}
Recent research has shown that LLMs significantly improve software engineering efficiency in various tasks, including code generation \cite{guo2024deepseek}, bug detection\cite{ma2023adapting}, automated documentation\cite{luo2024repoagent}, and software maintenance\cite{wang2024software}. Studies have shown that models such as Codex\cite{chen2021evaluating} and GPT-4\cite{achiam2023gpt} can generate functionally correct code with minimal human intervention, reducing development time\cite{hou2024large}. 
Furthermore, LLMs have improved software debugging by identifying and explaining code vulnerabilities with high accuracy\cite{shestov2025finetuning}. 
In requirements engineering, LLMs facilitate natural language processing for requirement elicitation and refinement, enhancing clarity and consistency\cite{vogelsang2024using}. 
Another key contribution is the acceleration of code review and refactoring processes, where LLMs assist in identifying anti-patterns and suggesting optimized implementations\cite{zhong2024can}. 

Besides, improving human-AI collaboration in software development attracts more attention, specifically integrating LLMs into requirements engineering.
Current LLM applications in requirements engineering primarily focus on single-task automation, such as requirement elicitation\cite{ataei2024elicitron}, specification\cite{ma2024specgen} and verification, while emerging multi-agent systems are advancing toward comprehensive multi-phase automation to reduce manual efforts\cite{liu2024large}.

% Despite these advancements, challenges remain in ensuring model reliability, mitigating bias, and integrating LLM-based tools into existing software engineering workflows. 

% Future research focuses on fine-tuning LLMs for domain-specific applications and improving human-AI collaboration in software development.

\subsection{LLM-based Agent}
% 待修改
LLM-based agents are autonomous systems that integrate reasoning, decision-making, and environmental interaction, and go beyond standalone LLMs by implementing multi-step task execution, long-term memory, and dynamic adaptive feedback\cite{xi2025rise}. 
From software engineering perspective, LLM-based agents exhibit multidimensional functionality, enabling both atomic task and end-to-end software development process involving multiple software engineering activities\cite{liu2024large}.
This holds equally true for requirements engineering. LLM-driven agents are not limited to isolated phases, such as requirements elicitation and modeling, but can also address tasks spanning the entire requirements engineering lifecycle.
Nevertheless, within Agile development frameworks, software projects increasingly de-emphasize producing traditional requirement artifacts, such as interpretable requirement models, formal specifications, and validation\cite{bhattacharya2024mastering}.
Instead, there are growing preference for approaches that deliver immediate benefits through direct mapping of software goals to user stories, with agent-driven techniques for automated story generation\cite{manish2024autonomous} and quality verification\cite{zhang2024llm, li2024simac} attracting particular research attention.

% MCP
\section{Threats to Validity}

In the human alignment experiments, we primarily evaluated the semi-automatic dataset construction method, as well as automated assessment methods for FHR and QuACE. During the experiments, each member of the human alignment team received questionnaires assessing the consistency of various user stories, alongside quality evaluations using the QuACE metric. For each case, participants were provided in advance with relevant project background and problem context, facilitating accurate judgment.

In our final experiments, we did not utilize a purely super-agent-based approach; instead, we employed a super-agent combined with the CoT method as our baseline. Preliminary experiments demonstrated that the pure Super-Agent approach performed inadequately regarding FHR and QuACE outcomes and sometimes failed to output results in the required format. Therefore, we considered the super-agent combined with CoT approach more suitable as a baseline. Additionally, the reasoning process of the Super-Agent was structured explicitly within the Impact Mapping framework, highlighting Goal2Story’s superior performance in goal-driven requirements elicitation tasks.
\section{Conclusions}

We propose Goal2Story, a multi-agent fleet for goal-driven RE that exclusively utilizes privately deployed sLLMs. By incorporating impact mapping as a core methodology, Goal2Story enhances RE performance on FHR and QuACE by decomposing complex and large-scale tasks into smaller, well-defined subtasks. To support evaluation and future applications, we introduce the StorySeek dataset, which captures key elements of the goal-driven RE process and presents a semi-automatic dataset construction method. Additionally, our work explores the identification of latent needs and demonstrates the potential of agent-based approaches for addressing this challenge in future research.

%%
% Acknowledgments.
\begin{acks}
Thanks to the anonymous reviewers for their helpful comments.
\end{acks}

\bibliographystyle{unsrt}
\bibliography{bibliography}

%%
%% Appendix, this is the place to put it.
\appendix
\section{Quality And Consistency Evaluation}
\label{app:QuACE_app}
\textbf{Quality And Consistency Evaluation (QuACE)}
\begin{itemize}
    \item Syntactic: a user story must include elements of actor (can be a person, a group, or an organization), action, and expected\_outcome, and contain no extra comments.
    \item Semantic: a user story must define the intended function and impact, focus on the problem, and avoid any ambiguous terms.
    \item Pragmatic: a user story must be a complete sentence that can be estimated for planning.
    \item Consistency: (1) Goal-related: a user story must be logically aligned with the goal, please remember that the relatedness could be indirect. (2) Factual-related: a user story must be consistent with the context, please remember that the factual consistency could be indirect. (Factual-related criteria is only used in the StorySeek dataset check)
\end{itemize}

\section{Human Alignment Experiment Details} \label{app:alignment}
% 设置全局表格间距
\setlength{\textfloatsep}{6pt}  % 默认 20pt，减小表格间距
\setlength{\intextsep}{6pt}      % 控制浮动对象（table, figure）的间距
\captionsetup{aboveskip=3pt, belowskip=3pt} % 控制 caption 的间距

\begin{table}[H]
    \centering
    \begin{tabular}{|l|c|c|c|}
        \hline
        \multirow{2}{*}{Tester} & \multicolumn{3}{c|}{Consistent Number} \\ \cline{2-4} 
                                & All              & Two             & None         \\ \hline
        Total                   & 17/ 37.78\%      & 19 / 42.22\%    & 9 / 20\%     \\ \hline
    \end{tabular}
    \caption{Cross-Model Consistency Validation Results for StorySeek}
    \label{tab:summary}
\end{table}

\begin{table}[H]
    \centering
    \begin{tabular}{|c|c|c|}
        \hline
        \multirow{2}{*}{\textbf{Predicted Results}} & \multicolumn{2}{c|}{\textbf{Ground Truth}}  \\
        \cline{2-3}
        & \textbf{hit} & \textbf{no-hit} \\
        \hline
        \textbf{hit} & 23 & 8 \\
        \cline{1-3}
        \textbf{no-hit} & 7 & 22 \\
        \hline
    \end{tabular}

    % ------ 第二部分: Precision / Recall / F1 / Alignment rate / FPR ------
    \begin{tabular}{|c|c|c|c|c|}
        \hline
        \textbf{Precision} & \textbf{Recall} & \textbf{F1} & \begin{tabular}{@{}c@{}}\textbf{Alignment}\\\textbf{rate}\end{tabular} & \textbf{FPR} \\
        \hline
        76.67\% & 74.19\% & 75.41\% & 75\% & 24.14\% \\
        \hline
    \end{tabular}

    \caption{FHR Human Alignment Experimental Results}
    \label{tab:total}
\end{table}

\begin{table}[H]
    \centering
    % ------ 第一部分: Ground Truth / Predicted Result ------
    \begin{tabular}{|c|c|c|}
        \hline
        \multirow{2}{*}{\textbf{Predicted Results}} & \multicolumn{2}{c|}{\textbf{Ground Truth}}
        \\ \cline{2-3}
        & \textbf{pass} & \textbf{failure} \\ 
        \hline
        \textbf{pass} & 33 & 10 \\ \cline{1-3}
        \textbf{failure} & 12 & 5 \\ 
        \hline
    \end{tabular}

    % ------ 第二部分: Precision / Recall / F1 / Alignment rate / FPR ------
    \begin{tabular}{|c|c|c|c|c|}
        \hline
        \textbf{Precision} & \textbf{Recall} & \textbf{F1} & \begin{tabular}{@{}c@{}}\textbf{Alignment} \\ \textbf{rate}\end{tabular} & \textbf{FPR} \\
        \hline
        73.33\% & 76.74\% & 75\% & 63.33\% & 70.59\% \\ 
        \hline
    \end{tabular}

    \caption{QuACE Human Alignment Experimental Results}
    \label{tab:another}
\end{table}

\end{document}